\documentclass{article}

\usepackage[english]{babel}

\usepackage[a4paper]{geometry}

\usepackage{amsmath}
\usepackage{graphicx}
\usepackage[colorlinks=true, allcolors=blue]{hyperref}
\usepackage{siunitx}
\usepackage{textcomp}
\usepackage{float}
\usepackage{xcolor}
\usepackage{a4wide}
\usepackage{tabularx} 
\usepackage{hyperref}
\usepackage{authblk}

\newcolumntype{Y}{>{\centering\arraybackslash}X}

\title{An Open Dataset for Temperature Modelling in Machine Tools}
\author[1,+,*]{C. Coelho}
\author[2,+]{D. Fernández}
\author[1,+]{M. Hohmann}
\author[2]{L. Penter}
\author[2]{S. Ihlenfeldt}
\author[1]{O. Niggemann}
\affil[1]{Institute for Artificial Intelligence, Helmut Schmidt University, 22043 Hamburg, Germany}
\affil[2]{Chair of Machine Tools Development and Adaptive Controls, Dresden University of Technology TUD, 01069 Dresden, Germany}
\affil[*]{Corresponding author: cecilia.coelho@hsu-hh.de}
\affil[+]{Shared co-first authorship.}
\date{}

\begin{document}
\maketitle

\begin{abstract}
This data set descriptor introduces a structured, high-resolution dataset of transient thermal simulations for a vertical axis of a machine tool test rig. The data set includes temperature and heat flux values recorded at 29 probe locations at 1800 time steps, sampled every second over a 30-minute range, across 17 simulation runs derived from a fractional factorial design. First, a computer-aided design model was de-featured, segmented, and optimized, followed by finite element (FE) modelling. Detailed information on material, mesh, and boundary conditions is included. To support research and model development, the dataset provides summary statistics, thermal evolution plots, correlation matrix analyses, and a reproducible Jupyter notebook. The data set is designed to support machine learning and deep learning applications in thermal modelling for prediction, correction, and compensation of thermally induced deviations in mechanical systems, and aims to support researchers without FE expertise by providing ready-to-use simulation data.


\end{abstract}



\section{Background \& Summary}



Thermal effects are widely recognized as the predominant source of machining inaccuracies, with thermally induced deformations contributing up to 70\% of the total geometric errors in machine tools \cite{Mayr2012}. Thus, accurately predicting temperature fields over time is crucial to achieving precision, reliability, and reduced production costs.
Traditional approaches relied heavily on manual compensation and simplistic regulation of internal heat sources. Despite significant advancements in thermal compensation and temperature control strategies \cite{Li2015, Feng2015, Fujishima2019}, achieving consistent, high-precision machining remains challenging. The complex thermo-mechanical behaviour of machine structures continues to compromise positional accuracy and process stability \cite{Mayr2012, Li2015}. Thermally induced displacements at the tool centre point (TCP) stem from a combination of internal heat sources, such as spindles, drive motors, and guideways, and external thermal disturbances, including ambient temperature fluctuations, acting along the thermo-elastic functional chain \cite{Feng2015, Fujishima2019}.

Current existing thermal modelling and thermal error correction models can be classified into four categories: statistical correlation models \cite{liu2020thermal,lee2002statistical}; transfer function-based models \cite{mares2013robustness,marevs2019strategy}; analytical and numerical structural models \cite{thiem2019online,thiem2014structural}; and, less explored, Artificial Intelligence-driven approaches \cite{mu2025review,zhang2012machine},



High-quality, publicly and ready available datasets for manufacturing processes, involving thermal behaviour in machine tools, are scarce. The main reason for this scarcity is that generating realistic, high-fidelity thermal data requires expertise in Finite Element (FE) modelling, including knowledge of geometry preparation, meshing, material properties, and boundary conditions. As a result, such data is typically out of reach for researchers in Machine Learning (ML) and Deep Learning (DL) who may lack expertise in these tools or processes.
To address these limitations, we present a structured, high-resolution dataset of transient thermal simulations designed to support the development of ML and DL surrogates \cite{brunton2022data}. Training surrogate models on this data significantly reduces the computational cost of predicting temperature and heat flux evolution under varying input conditions \cite{pavlivcek2019applicability,karandikar2015cost}. These models can then be used to optimise and perform error correction on demand, thus reducing production downtime and improving process efficiency.


The presented dataset captures temperature and heat flux at 29 probe node locations over time  of a vertical axis of a machine tool test rig. Seventeen distinct simulation runs with different initial conditions were derived from a $2^{(5-1)}$ fractional factorial design. These simulations represent a wide range of thermal conditions, making the dataset suitable for studying and modelling both specific (a single initial condition) or global behaviour. A total of 35 files are provided, comprising fixed geometry and material property data, and transient temperature and heat flux values. Each simulation run contains 1,800 time steps, resulting in values per second over a period of 30 minutes. Statistical analysis, time evolution plots, and correlation matrices are also included to offer insights into the data structure and variability. The statistical analysis is reproducible using the Jupyter notebook included in the repository, allowing researchers to explore, extend, or adapt the analysis pipeline with minimal setup.
The dataset is designed for the application of ML and DL models, supporting regression tasks, such as predicting thermal behaviour, and classification tasks, such as identifying component type or location. This work aims to accelerate the development of thermal surrogate models to enable wider adoption and development of ML and DL techniques in thermal modelling for machine tools.

\section{Methods}

\subsection*{Model Construction}

\begin{figure}[h]
    \centering
    \includegraphics[width=1\linewidth]{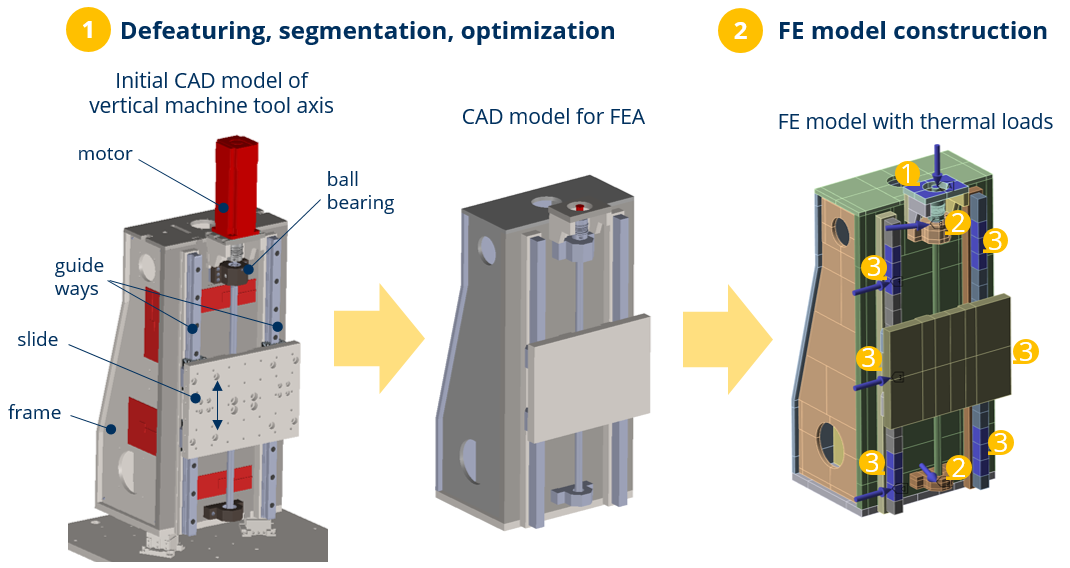}
    \caption{Modelling workflow: from CAD preparation to set thermal loads (with 1 the motor, 2 the bearings and 3 the guideways).}
    \label{fig:modeling-workflow}
\end{figure}

\begin{figure}
    \centering
    \includegraphics[width=0.85\linewidth]{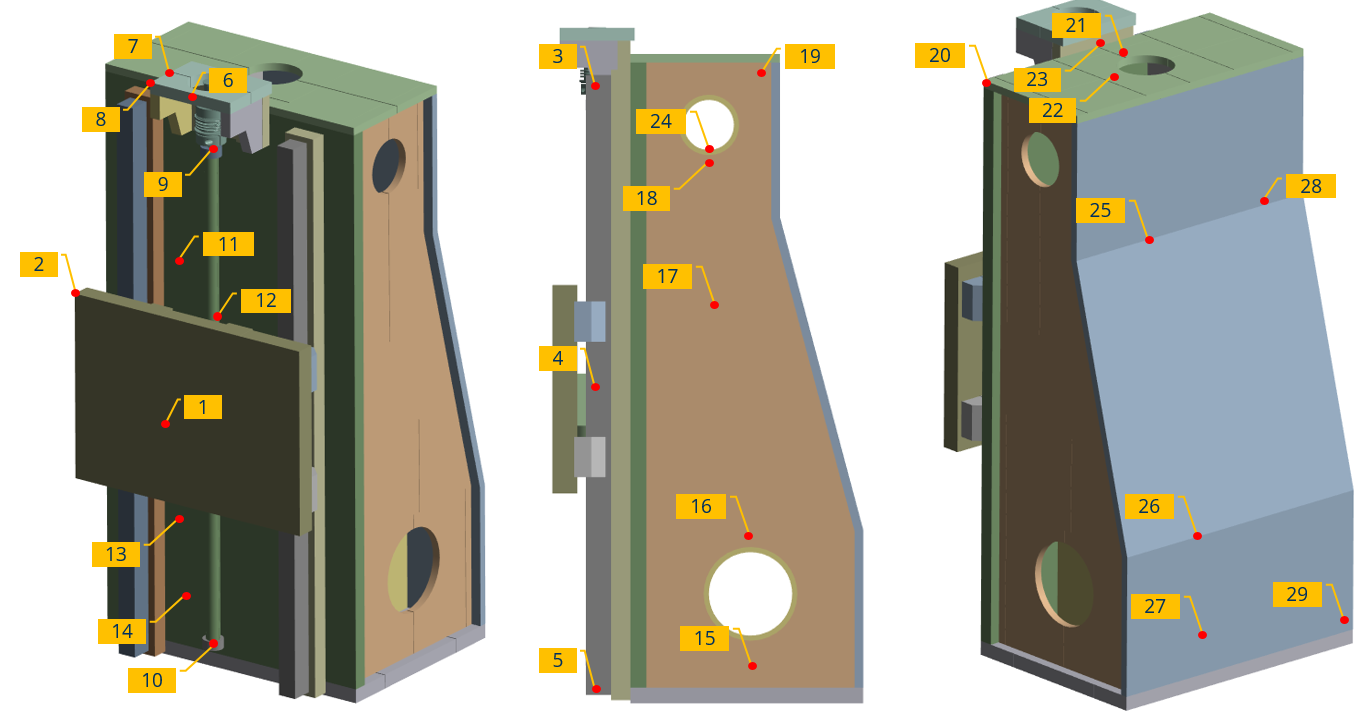}
    \caption{Sensor locations.}
    \label{fig:sensor_location}
\end{figure}

The FE model was generated by defeaturing, segmenting, and optimizing the initial Computer-Aided Design (CAD) assembly using Solidworks \cite{Pöhlmann2023}. Special emphasis was placed on the most important heat source components, such as the motor, bearings, and guideways. The node numbers were set according to a criteria that maximizes the quality of the thermal behaviour representation in the machine tool, achieved through finer meshing in regions with steep temperature gradients, such as heat sources and boundaries, while avoiding prohibitive computational costs. This approach incorporates a convergence study to ensure result stability, maintains optimal element aspect ratios to enhance accuracy, and balances precision with computational efficiency \cite{Inigo2024}.
Their locations are as follows, Fig. \ref{fig:sensor_location}: 1-2 (carrier), 3-5 (guideway), 6-8 (motor base), 9-10 (bearing), 11-14 (front structure), 15-19 (lateral structure), 20-23 (top structure), and 24-29 (back structure). Non-critical details, such as small holes and screws, were removed to simplify the meshing process and due to its non-relevant effect on the output of the current approach. The FE model was built using ANSYS. A mesh with 26,434 hexahedral elements and 157,327 nodes at a \qty{0.01}{\metre} resolution was created. Fig. \ref{fig:modeling-workflow} shows the FE CAD model preparation and the heat fluxes (heat loads) applied as part of the set of boundary conditions. 

\subsection*{Boundary Conditions}

The thermal boundary conditions were applied as follows: All exposed exterior surfaces, such as the motor and the guideways, are subject to convective heat transfer with the ambient air. The variable associated for convective effects was the air film coefficient, which represents the behaviour of a laminar or turbulent flow on the involved surfaces. Radiation was not considered in this study.
The set of boundary conditions also considered ground temperature, air temperature, initial machine temperature, and heat fluxes as thermal loads.
Regarding heat fluxes, only three main loads were considered and thus also simplified due to their complexity in real life. To simplify the approach, effects such as velocity-dependent friction or non-linearities due to changes in material properties were avoided.
From the previous, the thermal loads were considered constant on the surfaces implied related to the main heat sources: the motor, bearings, and guideways. Each considering different magnitudes based on their relevance in the heat contribution (see Table \ref{tab:heat_flux}). Motor and bearings received uniform heat fluxes, while the guideways were split into three thermal load locations. The top and bottom segments were assumed to have the same heat flux magnitude, while the middle segment had a higher magnitude due to its longer exposure to the carrier's  movement  along its path. Fig.\ref{fig:modeling-workflow} on the far right shows the locations of the heat fluxes, numbered 1 for the motor, 2 for the bearings and 3 for the guideways.

    \begin{table}[h]
        \centering
         \caption{Heat flux magnitudes.}
         \label{tab:heat_flux}
          \begin{tabular}{ccc}
         \hline
          \textbf{Source} & \textbf{Heat flux} & \textbf{Unit} \\
         \hline
          Motor & 1000  & W/m² \\
          Bearings (top and bottom) & 300 & W/m²\\
         Guideways: top and bottom & 200 & W/m²\\
         Guideways: middle & 600 & W/m²\\
          \hline
         \end{tabular}
    \end{table}      

Then, the Design of Experiment (DoE) was carried out with five factors and a two-level factorial design with one centre point, resulting in 17 non-repetitive runs due to the nature of the data generation. 
Table \ref{tab:DoE} summarises the thermal boundary conditions used across the simulation runs: the heat fluxes used for each run and component was obtained by multiplying the heat flux magnitudes presented in Table \ref{tab:heat_flux} with the Heat Flux Level factor. The air film coefficient represents the convective heat transfer and models the effectiveness of passive cooling. The ambient temperature defines the temperature of the direct machine environment, which influences heat dissipation. The system initial temperature defines the uniform starting temperature of the machine. The ground temperature represents the thermal boundary condition at the base of the machine tool and simulates heat exchange with the floor or mounting surface.

\begin{table}[h]
    \centering
    \caption{Design of experiments thermal boundary conditions for the 17 runs.}
    \label{tab:DoE}
    \begin{tabularx}{\textwidth}{lYYYYY} 
        \hline
        \textbf{Run} & \textbf{Heat Flux Level, $\mathrm{W}/\mathrm{m}^{2}$} & \textbf{Air Film Coefficient, $\mathrm{W}/\mathrm{m}^{2}\cdot^\circ\mathrm{C}$ } & \textbf{Ambient Temperature, $^\circ\mathrm{C}$} & \textbf{System Initial Temperature, $^\circ\mathrm{C}$} & \textbf{Ground Temperature, $^\circ\mathrm{C}$} \\
        \hline
        1 & 0.5 & 10 & 20 & 20 & 18 \\
        2 & 1.5 & 10 & 20 & 20& 38 \\
        3 & 0.5 & 50 & 20 &20 & 38\\
        4 & 1.5 & 50 & 20 & 20& 18\\
        5 & 0.5 & 10 & 40 & 20& 38\\
        6 & 1.5 & 10 & 40 &20 & 18\\
        7 & 0.5 & 50 & 40 & 20& 18\\
        8 & 1.5 & 50 & 40 & 20 & 38\\
        9 & 0.5 & 10 & 20 & 40 & 38\\
        10 & 1.5 & 10 & 20 &40 & 18\\
        11 & 0.5 & 50 & 20 & 40& 18\\
        12 & 1.5 & 50 & 20 & 40& 38\\
        13 & 0.5 & 10 & 40 &40 & 18\\
        14 & 1.5 & 10 & 40 &40 & 38\\
        15 & 0.5 & 50 & 40 & 40& 38\\
        16 & 1.5  & 50 & 40 & 40& 18\\
        17 & 1.0 & 30 & 30 & 30 & 28 \\
        \hline
    \end{tabularx}
\end{table}

Each transient simulation run was solved for a period of 30 minutes $(t=\qty{1200}{s})$, with outputs recorded every second.

\section{Data Records}

The dataset consists of 35 data files in standard Comma-Separated Values (CSV):
\begin{itemize}
    \item \textbf{Mesh and properties:} A single file (\emph{TransientThermalSimulationFE\_MaterialProperties\_*\footnote{\texttt{*} is used as a wildcard (regex syntax) indicating there exists more characters after the base name but are not relevant for main file type identification (run date, mesh size).}.txt}), common to the 17 simulation runs, listing the FE model node identity number and coordinates, and material properties (density and thermal conductivity), Table \ref{tab:meshFile};

        \begin{table}[H]
        \caption{Format of the mesh and properties file.}
        \resizebox{\textwidth}{!}{
        \begin{tabular}{cccccc}
        \hline
        \textbf{Node Number} & \textbf{X Position} & \textbf{Y Position} & \textbf{Z Position} & \textbf{Density} & \textbf{Thermal Conductivity} \\ \hline
        1                    & 0.064100               & -0.133366              & 0.035898               & 7850                    & 60.5                                 \\
        \dots                &          \dots                &           \dots               &        \dots                  &           \dots                &             \dots                           \\
        29                  & 0.064100               & -0.153729              & 0.073535               & 7850                    & 60.5                                 \\ \hline
        \end{tabular}
        }
        \label{tab:meshFile}
        \end{table}

    \item \textbf{Temperature fields:} 17 files (\emph{TransientThermalSimulationFE\_RunN\_Temperature\_*.txt}, with $\text{N} \in 1, \dots, 17$), each corresponding to one simulation run, Table \ref{tab:tempFile}. Columns represent: steps; time in seconds; temperature at each of the 29 probe nodes in $^\circ\mathrm{C}$;

    \item \textbf{Heat flux fields:} 17 files (\emph{TransientThermalSimulationFE\_RunN\_HeatFlux\_*.txt}, with $\text{N} \in 1, \dots, 17$), each corresponding to one simulation run, Table \ref{tab:heatFile}. Columns represent: steps; time in seconds; heat flux at each of the 29 probe nodes (in $\mathrm{W}/\mathrm{m}^{2}$).
\end{itemize}

        \begin{table}[H]
        \centering
        \begin{minipage}{0.48\textwidth}
            \centering
            \caption{Format of a temperature field file.}
            \resizebox{\textwidth}{!}{
            \begin{tabular}{ccccc}
                \hline
                \textbf{Steps} & \textbf{Time} & \textbf{Probe1} & \textbf{$\dots$} & \textbf{Probe29}  \\ \hline
                1 & 1 & 20 & $\dots$ & 19.997 \\
                $\dots$ & $\dots$ & $\dots$ & $\dots$ & $\dots$ \\
                1800 & 1 & 20.269 & $\dots$ & 18.378 \\ \hline
            \end{tabular}
            }
            \label{tab:tempFile}
        \end{minipage}
        \hfill
        \begin{minipage}{0.48\textwidth}
            \centering
            \caption{Format of a heat flux field file.}
            \resizebox{\textwidth}{!}{
            \begin{tabular}{ccccc}
                \hline
                \textbf{Steps} & \textbf{Time} & \textbf{Probe1} & \textbf{$\dots$} & \textbf{Probe29}  \\ \hline
                1 & 1 & 4.7853e-5 & $\dots$ & 2.2382e-8 \\
                $\dots$ & $\dots$ & $\dots$ & $\dots$ & $\dots$ \\
                1800 & 1 & 6.0656 & $\dots$ & 346.11 \\ \hline
            \end{tabular}
            }
            \label{tab:heatFile}
        \end{minipage}
        \end{table}

All data files are openly available at \href{https://github.com/CeciliaCoelho/temperatureModellingDataset}{github.com/temperatureModellingDataset}. The units and data organisation are consistent across files and the column headers and file naming are straightforward to identify the simulation run and variable.

\subsection*{Statistical Analysis and Data Visualisation}


To complement the dataset, a statistical analysis and visualisation is provided.

The fixed material and mesh data include 29 probe nodes sampled from the FE model. All nodes are located in components that share the same material, structural steel with a density of 7850 $\mathrm{Kg}/\mathrm{m}^{3}$ and thermal conductivity of 60.5 $\mathrm{W}/\mathrm{m}^{2}$. The node coordinates basic statistics, minimum (min) and maximum (max) value, mean and standard deviation (std) are organised in Table \ref{tab:nodeStats}.

        \begin{table}[H]
        \caption{Node coordinate statistics.}
        \centering
        \resizebox{0.6\textwidth}{!}{
        \begin{tabular}{cccccc}
        \hline
        \textbf{Coordinate}       & \textbf{Min} & \textbf{Max} & \textbf{Mean} & \textbf{Std} & \textbf{Unit} \\ \hline
                    x   &          0.0641 &        0.0641      &        0.0641      &    0   & $\mathrm{m}$ \\ 
                    y   &         -0.2058  &        -0.1242      &   -0.1664           &  0.0307     & $\mathrm{m}$ \\ 
                    z   &          0.0248 &      0.0952        &     0.0547         &   0.0260    & $\mathrm{m}$ \\ 
                    \hline
        \end{tabular}
        }
        \label{tab:nodeStats}
        \end{table}

Each of the 17 temperature field files contains 1800 data points recorded at 29 probe nodes. Across all simulation runs, the basic statistics are organised in Table \ref{tab:tempStats}.

        \begin{table}[H]
        \caption{Temperature field statistics.}
        \centering
        \resizebox{0.5\textwidth}{!}{
        \begin{tabular}{cccccc}
        \hline
        \textbf{Node} & \textbf{Min} & \textbf{Max} & \textbf{Mean} & \textbf{Std}  & \textbf{Unit} \\ \hline
        1   & 20.00 & 40.84 & 30.15 & 7.96 &  $^\circ\mathrm{C}$ \\
        2   & 20.00 & 41.93 & 30.54 & 7.89 &  $^\circ\mathrm{C}$ \\
        3   & 20.00 & 41.70 & 30.47 & 7.89 &  $^\circ\mathrm{C}$ \\
        4   & 20.04 & 44.34 & 32.07 & 8.00 &  $^\circ\mathrm{C}$ \\
        5   & 19.35 & 40.21 & 29.83 & 6.49 &  $^\circ\mathrm{C}$ \\
        6   & 20.07 & 60.99 & 39.19 & 9.92 &  $^\circ\mathrm{C}$ \\
        7   & 20.07 & 54.82 & 36.13 & 8.85 &  $^\circ\mathrm{C}$ \\
        8   & 20.07 & 57.14 & 37.03 & 9.13 &  $^\circ\mathrm{C}$ \\
        9   & 20.00 & 41.31 & 30.28 & 8.04 &  $^\circ\mathrm{C}$ \\
        10   & 20.00 & 40.79 & 30.13 & 8.01 &  $^\circ\mathrm{C}$ \\
        11   & 19.81 & 40.51 & 30.08 & 7.27 &  $^\circ\mathrm{C}$ \\
        12   & 18.41 & 40.00 & 28.62 & 7.63 &  $^\circ\mathrm{C}$ \\
        13   & 18.25 & 40.00 & 28.54 & 7.73 &  $^\circ\mathrm{C}$ \\
        14   & 19.85 & 40.00 & 29.98 & 7.81 &  $^\circ\mathrm{C}$ \\
        15   & 20.00 & 40.38 & 30.06 & 7.94 &  $^\circ\mathrm{C}$ \\
        16   & 20.00 & 40.18 & 30.02 & 7.94 &  $^\circ\mathrm{C}$ \\
        17   & 20.00 & 40.94 & 30.17 & 7.90 &  $^\circ\mathrm{C}$ \\
        18   & 20.00 & 41.06 & 30.20 & 7.93 &  $^\circ\mathrm{C}$ \\
        19   & 20.00 & 43.50 & 30.87 & 7.91 &  $^\circ\mathrm{C}$ \\
        20   & 20.00 & 46.45 & 32.16 & 8.11 &  $^\circ\mathrm{C}$ \\
        21   & 20.00 & 41.02 & 30.20 & 8.09 &  $^\circ\mathrm{C}$ \\
        22   & 20.00 & 40.07 & 30.01 & 7.88 &  $^\circ\mathrm{C}$ \\
        23   & 19.39 & 40.00 & 29.72 & 7.07 &  $^\circ\mathrm{C}$ \\
        24   & 18.64 & 40.00 & 28.99 & 6.62 &  $^\circ\mathrm{C}$ \\
        25   & 20.00 & 40.08 & 30.01 & 7.89 &  $^\circ\mathrm{C}$ \\
        26   & 18.38 & 40.00 & 28.64 & 7.28 &  $^\circ\mathrm{C}$ \\
        27   & 20.00 & 42.09 & 30.48 & 7.85 &  $^\circ\mathrm{C}$ \\
        28   & 19.29 & 40.13 & 29.78 & 6.82 &  $^\circ\mathrm{C}$ \\
        29   & 20.00 & 40.49 & 30.09 & 7.95 &  $^\circ\mathrm{C}$ \\
                    \hline
        \end{tabular}
        }
        \label{tab:tempStats}
        \end{table}

\begin{figure}[H]
    \centering
    \includegraphics[width=0.6\linewidth]{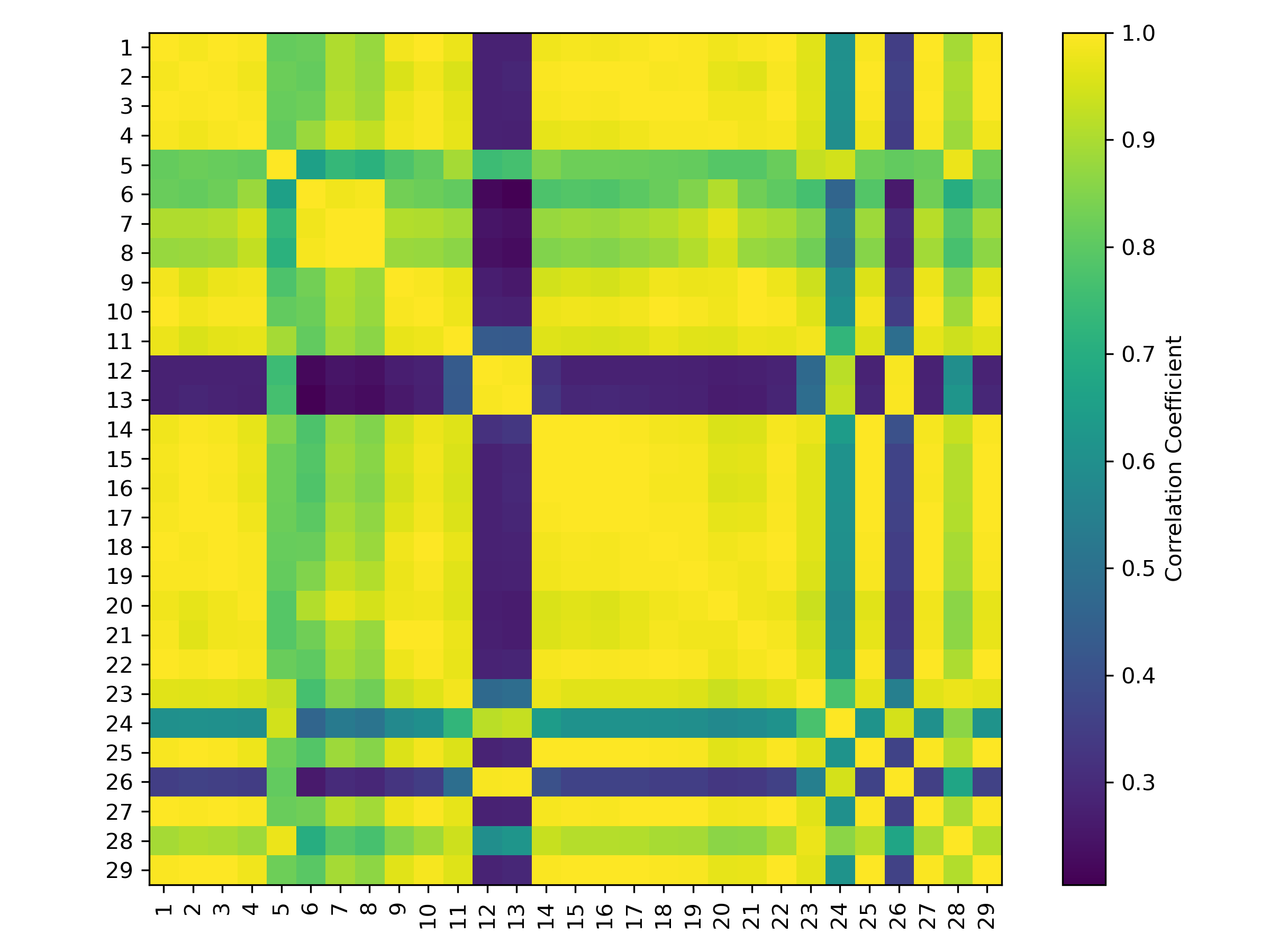}
    \caption{Correlation matrix for the temperature field data.}
    \label{fig:corrTemp}
\end{figure}

To detect and analyse possible linear relationships between thermal responses at different probe node locations, the correlation matrix, using the Pearson coefficient, across all 17 runs was computed, Fig. \ref{fig:corrTemp}.

Most of the nodes show high correlation between each other, indicating a uniform temperature field in the system. Only a few nodes in the front structure (12, 13) and back structure (26) have a lower correlation with other locations. In the context of ML or DL applications, highly correlated structures can be excluded to minimise redundancy and reduce computational costs during models training and testing \cite{bishop2006pattern}.

\begin{figure}[H]
    \centering
    \includegraphics[width=0.9\linewidth]{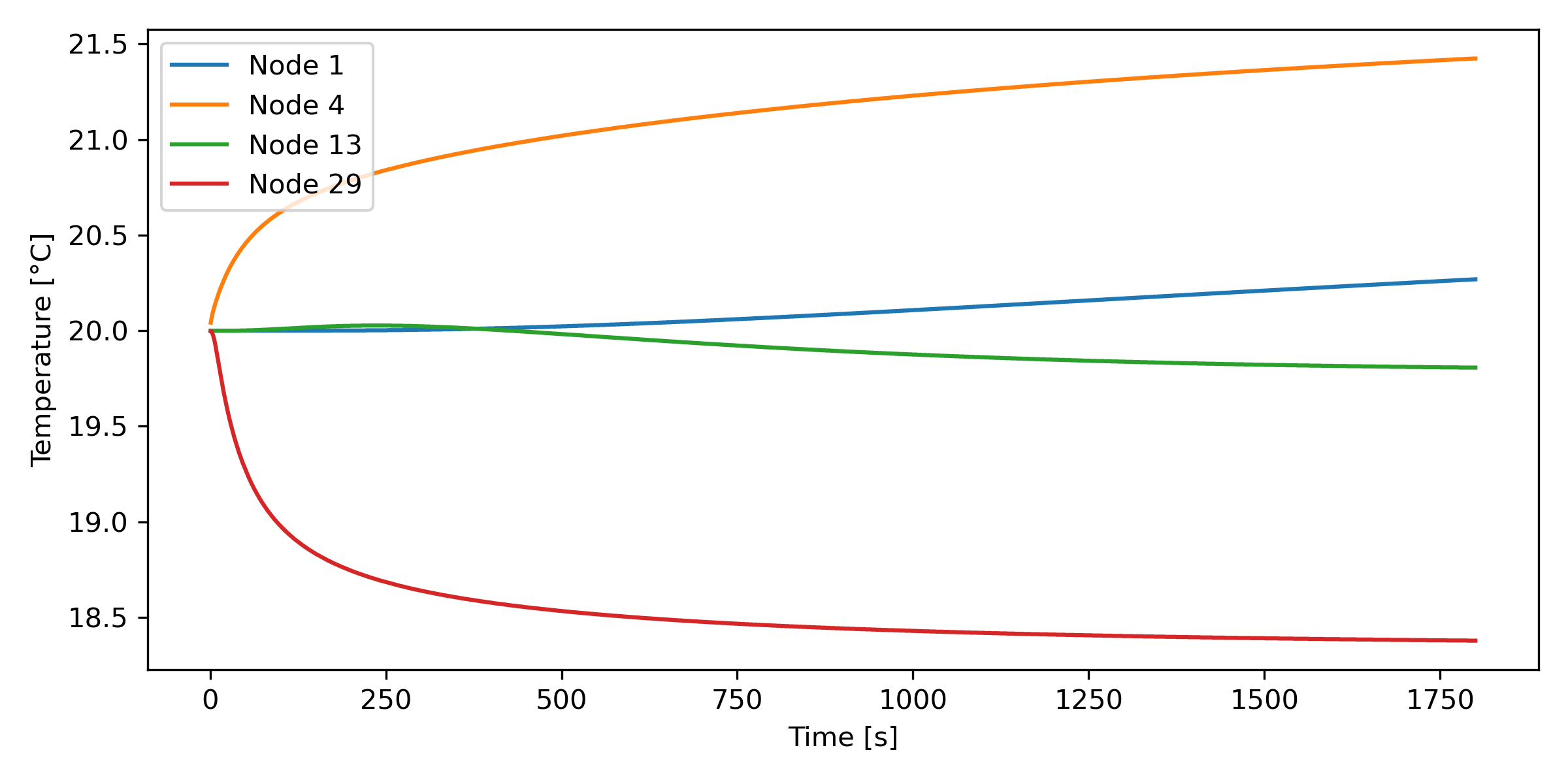}
    \caption{Temperature curve for selected node 1, 4, 13, and 28.}
    \label{fig:plotTemp}
\end{figure}

Furthermore, plots of the temperature increment in transient state at selected nodes are provided in Fig. \ref{fig:plotTemp}.

Similarly, each 17 heat flux field files contain 1800 data points recorded at 29 probe nodes. The main statistics, across all simulation runs, are organised in Table \ref{tab:heatStats}. 

        \begin{table}[H]
        \caption{Heat flux field statistics.}
        \centering
        \resizebox{0.6\textwidth}{!}{
        \begin{tabular}{cccccc}
        \hline
        \textbf{Node} & \textbf{Min} & \textbf{Max} & \textbf{Mean} & \textbf{Std} & \textbf{Unit} \\ \hline
        1   & 0.00 & 961.19 & 139.79 & 207.30 &  $\mathrm{W}/\mathrm{m}^{2}$ \\
        2   & 0.00 & 1574.90 & 193.82 & 270.92 &  $\mathrm{W}/\mathrm{m}^{2}$ \\
        3   & 0.02 & 1675.20 & 217.74 & 306.07 &  $\mathrm{W}/\mathrm{m}^{2}$ \\
        4   & 261.48 & 2068.10 & 851.71 & 403.88 &  $\mathrm{W}/\mathrm{m}^{2}$ \\
        5   & 0.35 & 1678.10 & 274.16 & 315.89 &  $\mathrm{W}/\mathrm{m}^{2}$ \\
        6   & 352.41 & 1499.60 & 999.28 & 484.84 &  $\mathrm{W}/\mathrm{m}^{2}$ \\
        7   & 347.81 & 2192.80 & 1223.93 & 609.48 &  $\mathrm{W}/\mathrm{m}^{2}$ \\
        8   & 339.66 & 1537.20 & 978.86 & 474.87 &  $\mathrm{W}/\mathrm{m}^{2}$ \\
        9   & 4.43 & 7669.20 & 2022.62 & 2048.50 &  $\mathrm{W}/\mathrm{m}^{2}$ \\
        10   & 14.48 & 1741.00 & 559.93 & 438.46 &  $\mathrm{W}/\mathrm{m}^{2}$ \\
        11   & 0.00 & 896.78 & 199.57 & 221.40 &  $\mathrm{W}/\mathrm{m}^{2}$ \\
        12   & 0.00 & 307.34 & 59.41 & 78.86 &  $\mathrm{W}/\mathrm{m}^{2}$ \\
        13   & 0.18 & 1013.80 & 474.39 & 319.01 &  $\mathrm{W}/\mathrm{m}^{2}$ \\
        14   & 15.45 & 10656.00 & 2539.30 & 2130.18 &  $\mathrm{W}/\mathrm{m}^{2}$ \\
        15   & 1.33 & 7990.10 & 1725.73 & 1641.09 &  $\mathrm{W}/\mathrm{m}^{2}$ \\
        16   & 0.00 & 971.77 & 162.68 & 187.43 &  $\mathrm{W}/\mathrm{m}^{2}$ \\
        17   & 0.00 & 972.21 & 140.09 & 173.57 &  $\mathrm{W}/\mathrm{m}^{2}$ \\
        18   & 0.00 & 993.88 & 152.76 & 194.35 &  $\mathrm{W}/\mathrm{m}^{2}$ \\
        19   & 0.00 & 993.95 & 107.68 & 169.81 &  $\mathrm{W}/\mathrm{m}^{2}$ \\
        20   & 0.00 & 1335.30 & 159.34 & 238.95 &  $\mathrm{W}/\mathrm{m}^{2}$ \\
        21   & 0.00 & 940.24 & 185.38 & 181.78 &  $\mathrm{W}/\mathrm{m}^{2}$ \\
        22   & 0.00 & 1319.50 & 188.18 & 246.72 &  $\mathrm{W}/\mathrm{m}^{2}$ \\
        23   & 0.02 & 3667.30 & 1728.13 & 998.05 &  $\mathrm{W}/\mathrm{m}^{2}$ \\
        24   & 0.00 & 732.85 & 206.60 & 204.02 &  $\mathrm{W}/\mathrm{m}^{2}$ \\
        25   & 0.00 & 978.31 & 236.59 & 282.97 &  $\mathrm{W}/\mathrm{m}^{2}$ \\
        26   & 0.00 & 2694.70 & 917.34 & 757.36 &  $\mathrm{W}/\mathrm{m}^{2}$ \\
        27   & 0.26 & 9130.60 & 3145.15 & 2538.94 &  $\mathrm{W}/\mathrm{m}^{2}$ \\
        28   & 0.00 & 1692.00 & 320.15 & 405.99 &  $\mathrm{W}/\mathrm{m}^{2}$ \\
        29   & 32.00 & 11828.00 & 2814.57 & 2425.87 &  $\mathrm{W}/\mathrm{m}^{2}$ \\
                    \hline
        \end{tabular}
        }
        \label{tab:heatStats}
        \end{table}

\begin{figure}[H]
    \centering
    \includegraphics[width=0.6\linewidth]{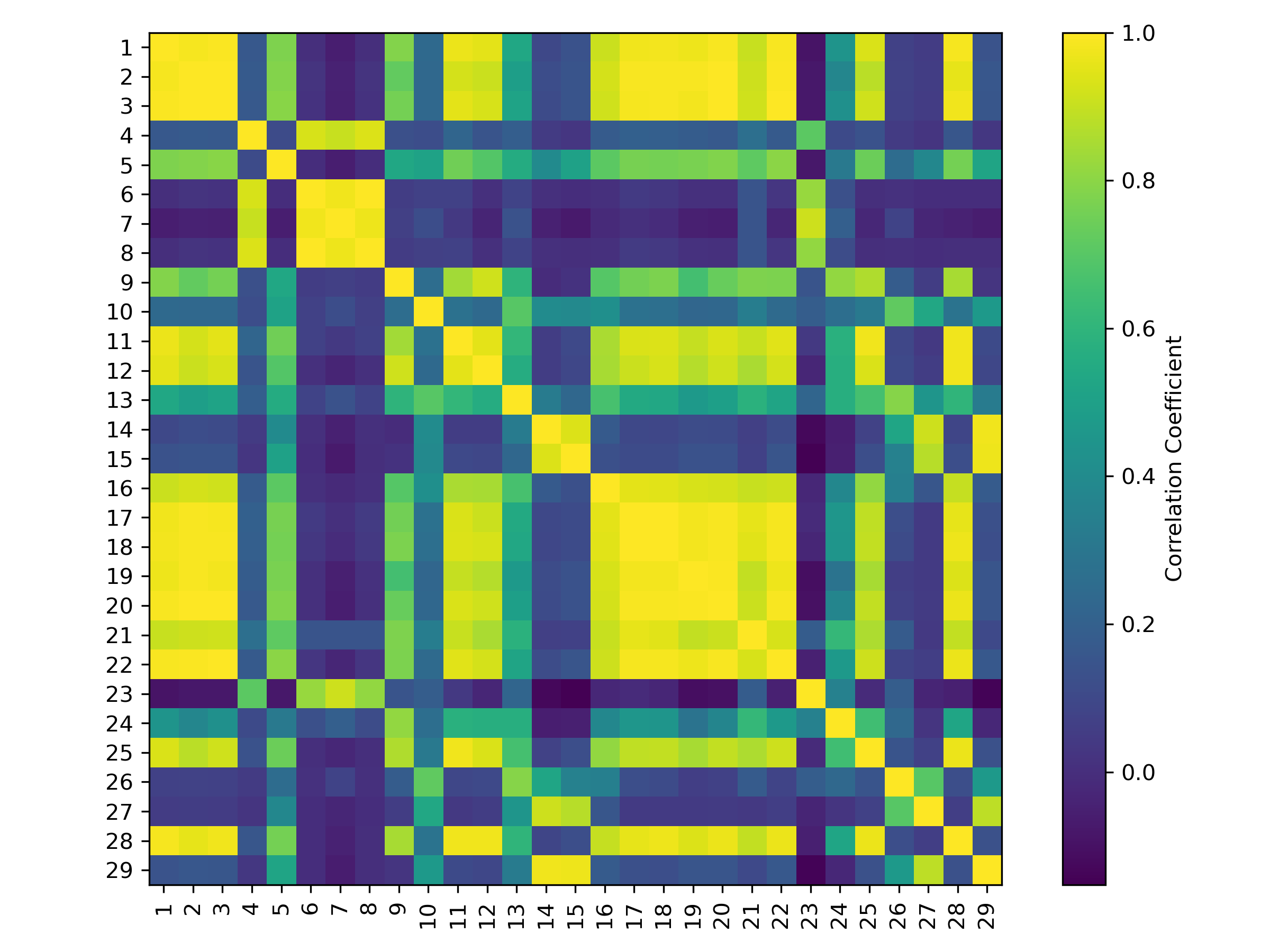}
    \caption{Correlation matrix for the heat flux data.}
    \label{fig:corrHeat}
\end{figure}

Additionally, the correlation matrix, using the Pearson coefficient, across all 17 runs was computed, Figure \ref{fig:corrHeat}.

The nodes on the carrier (1, 2, and 3) show a strong correlation, indicating similar heat flux behaviour. In contrast, the nodes in the motor base (6, 7, and 8) exhibit a strong internal correlation with the middle guideway (4) and are no correlation with other locations. In an ML or DL application, highly correlated structures can be removed to reduce training and testing computational costs \cite{bishop2006pattern}.

\section{Technical Validation}
The simulations were validated for internal consistency and physical plausibility following the methodology of Thiem et al. \cite{Thiem2023}. This process included de-featuring, segmenting functional components, and defining all surfaces that interact with thermal boundary conditions, followed by meshing. The meshing process was streamlined based on prior improvements, such as optimizing element size distribution to reduce computational overhead, implementing automated mesh refinement in high-gradient thermal zones, adopting a pre-defined convergence threshold to minimize manual iterations, and utilizing symmetry to halve the model domain where applicable.
Notably, the FE model was simplified by excluding nonlinear effects, such as friction and temperature-dependent material properties, as these would significantly increase computational complexity
First, repeated centre-point runs were found to produce identical temperature curves, confirming numerical stability. Additionally, the thermal responses behaved as expected across all runs: higher input fluxes or warmer ambient temperatures resulted in higher node temperatures. No negative, physically implausible, or outliers values were detected.

Although experimental measurements are not provided, comparable FE models have been validated in prior studies \cite{Thiem2023, Friedrich2024, Kumar2023}. Furthermore, the modelling workflow, Fig. \ref{fig:modeling-workflow}, is a common approach in the literature \cite{nicholson2008finite}.

\section{Usage Notes} 

The dataset and statistical analysis Jupyter notebook are available on the \emph{Github} platform at \href{https://github.com/CeciliaCoelho/temperatureModellingDataset}{github.com/temperatureModellingDataset}.

\subsection*{Application examples}

This dataset is intended to be used for training and testing ML or DL thermal surrogate models, aiming at reducing the computational cost of predicting the thermal behaviour of machine tool components. Potential users should note:

\begin{itemize}
    \item \textbf{Data format:} All data files are in CSV, being compatible with any programming or/and data analysis environment (\emph{Python}, \emph{R}, \emph{MATLAB}, etc.). Column headers are provided as the first row of each file and variable units are given in the International System of Units;

    \item \textbf{Probe nodes:} Information on 29 probe nodes, at key locations, is provided, comprising of 3-dimensional coordinates and material properties. 

    \item \textbf{Application examples:} The dataset can be used for both regression and classification tasks, in machine learning and deep learning \cite{bishop2006pattern}. For regression, the data can be seen as a time-series and used to model and predict the transient thermal behaviour (temperature and heat flux) at specific locations, given known input conditions. This allows the development of surrogate models (for digital twins, real-time monitoring, or rapid thermal design evaluation) capable of predicting the behaviour with lower computational cost than FE simulations. 
    
    For classification, the data can be used to classify the different components (bearings, motor, guideways) and their location (front, top, middle, centre, lateral, back). Based on their thermal behaviour over time, models can learn to detect structural roles or positional identifiers which is useful for automated model annotation, component tracking, or anomaly detection.
\end{itemize}

\subsection*{License}

The license for the dataset and all external material is Creative Commons Attribution 4.0 International. Researchers are free to share and adapt the presented data set, but they must give credit to the authors with a reference, provide a link to the license, and indicate changes to the original data set and other additional material.

\section{Data availability}

All data supporting this Data Descriptor are available at \href{https://github.com/CeciliaCoelho/temperatureModellingDataset}{github.com/CeciliaCoelho/temperatureModellingDataset} composing of: 17 heat flux and 17 temperature field files with values for 29 probe nodes and a sampling of one data point per second across 30 minutes; a properties file with node information and material properties for the 17 runs; a Jupyter notebook for data analysis. All resources are under license Creative Commons Attribution 4.0 International.

\section{Code availability}

No custom code is needed to access the data. The \emph{Python} code for basic statistical analysis and data visualisation can be found in a Jupyter notebook on \href{https://github.com/CeciliaCoelho/temperatureModellingDataset}{\emph{Github}}.

\section{Acknowledgements}

C. Coelho would like to thank the KIBIDZ project funded by dtec.bw---Digitalization and Technology Research Center of the Bundeswehr; dtec.bw is funded by the European Union---NextGenerationEU.
M. Hohmann is funded by the Deutsche Forschungsgemeinschaft (DFG, German Research Foundation) – 520460697.
D. Fernández is funded by the Deutsche Forschungsgemeinschaft (DFG, German Research Foundation)  - F-009765-551-521-1130701.







\subsection*{Contributions}

Conceptualization and Methodology: C.C., and D.F.; Software: D.F., C.C, and M.H.; Hardware: D.F., C.C, and M.H.; Validation: C.C., and D.F.; Formal analysis: D.F., C.C, and M.H.; Resources, O.N.; Data curation: C.C and M.H.; Writing—original draft preparation: D.F., C.C, and M.H.; Writing—review and editing, D.F., C.C, M.H., L.P.; Visualization: D.F., and M.H.; Supervision and Project Administration: C.C., O.N., L.P, S.I.; Funding acquisition: O.N., S.I. 
All authors reviewed the manuscript.

\section{Competing interests}

The authors declare no competing interests.


\end{document}